%% file: simp-STT-b.tex
\documentclass[fleqn,11pt]{article}
\usepackage{amsmath,amssymb,amsfonts,graphicx}

\input preamble.tex

\def\eqn#1{\eq\eqref{#1}}
\def\rf{\eqref}
\def\mn{_{\mu\nu}}
\def\MN{^{\mu\nu}}
\def\mN{_\mu^\nu}

\def\N{{\mathbb N}}
\def\R{{\mathbb R}}
\def\cF{{\cal F}}

\def\kappa{\varkappa}

\def\GR{general relativity}
\def\sph{spherically symmetric}
\def\ssph{static, spherically symmetric}

\def\bh{black hole}
\def\bhs{black holes}
\def\wh{wormhole}
\def\whs{wormholes}
 
\def\emag{electromagnetic}
\def\Scw{Schwarz\-schild}
\def\RN{Reiss\-ner-Nord\-str\"om}

\usepackage{color}

\begin{document}
\thispagestyle{empty}
\twocolumn[

\bigskip

\Title {Arbitrary static, spherically symmetric space-times \yy
		as solutions of scalar-tensor gravity}
	
\Aunames{Kirill A. Bronnikov,\au{a,b,c,1} Kodir Badalov,\au{d} and Rustam Ibadov\au{d,2}}

\Addresses{\small
\adr a {Center fo Gravitation and Fundamental Metrology, VNIIMS, 
		Ozyornaya ul. 46, Moscow 119361, Russia}
\adr b {	Institute of Gravitation and Cosmology, RUDN University, 
		ul. Miklukho-Maklaya 6, Moscow 117198, Russia}
\adr c 	{National Research Nuclear University ``MEPhI'', 
	Kashirskoe sh. 31, Moscow 115409, Russia}
\adr d {Department of Theoretical Physics and Computer Science, Samarkand State University, 
		Samarkand 140104, Uzbekistan}
		}

\bigskip

\Abstract
   {It is shown that an arbitrary static, spherically symmetric metric can be presented as an exact 
   solution of a scalar-tensor theory (STT) of gravity with certain nonminimal coupling function 
   $f(\phi)$ and potential $U(\phi)$. The scalar field in this representation can change its nature
   from canonical to phantom on certain coordinate spheres. This representation,
   however, is valid in general not in the full range of the radial coordinate but only piecewise. 
   Two examples of STT representations are discussed: for the \RN\ metric and for the 
   Simpson-Visser regularization of the \Scw\ metric (the so-called black bounce space-time).   
    }
\bigskip

] 
\email 1 {kb20@yandex.ru} 
\email 2{ibrustam@mail.ru}

\section{Introduction}

   Some recent studies in gravity theory consider examples of the space-time metric with properties 
   of interest that are not obtained as solutions to the field equations of \GR\ or its extensions.
   Many of them are nonsingular and are used as possible phenomenological descriptions of 
   space-times expected as a result of regularization due to quantum phenomena, both in the framework
   of semiclassical gravity and in various approaches to quantum gravity. Thus, for example,  
   Simpson and Visser  (SV) \cite{simp18} proposed a way to remove the singularity at $r=0$ in the
   \Scw\ metric by simply replacing the spherical radius $r$ with the expression 
   $r(u) = \sqrt{x^2 + b^2}$, where $x \in \R$ is a new radial coordinate, so that 
   the radius $r(x)$ does not reach the zero value.\footnote
   		{To avoid confusion, let us keep the notation $r$ for the spherical radius in \ssph\ space-times, 
  		such that $r^2 = -g_{\theta\theta}$, and denote other radial coordinates by other letters.}  
    As a result, we obtain the nonsingular metric
\bearr                    \label{ds-SV}
		ds^2 = \bigg(\! 1 - \!\frac{2m}{\sqrt{x^2\! +\! b^2}}\bigg)dt^2 
			- \bigg(\! 1 -\! \frac{2m}{\sqrt{x^2\! + \!b^2}}\bigg)^{-1}\! dx^2 
\nnnv \ \	
			- (x^2 + b^2)d\Omega^2,\quad\     d\Omega^2= d\theta^2 + \sin^2\theta d\varphi^2,
\ear  
  with a \Scw\ asymptotic behavior at large $|x|$ and mass $m$, but with a regular minimum of
  the radius, $r = b >0$ at $x=0$, instead of the singularity $r=0$. The geometry \rf{ds-SV} 
  represents a traversable  \wh\ with a throat at $x=0$ if $b > 2m$, a regular \bh\ with two 
  simple horizons at $x = \pm \sqrt{4m^2 -b^2}$ if $b < 2m$, or a regular \bh\ with a single 
  extremal horizon at $x=0$ if $b =2m$. 
  
  Similar regularizations were later obtained for the Reissner-Nordstr\"{o}m \cite{franzin21} 
  and other singular metrics \cite{lobo20}. The SV trick can be described more generally as replacing 
  a certain parameter of the metric, whose zero value corresponds to a singularity, with a manifestly
  nonzero quantity like  $\sqrt{u^2 + b^2}$, where $b > 0$ is a regularization parameter. 
  In particular, this looks as a simple way to represent possible effects of quantum gravity within a 
  classical framework, leaving aside any unknown quantization details. Besides, new geometries
  emerging in this way may be of interest by themselves.
  
  The resulting diverse and rich geometries have attracted rather much interest, and their 
  stationary rotating counterparts were also obtained and studied \cite{mazza21, xu21, shaikh21}, 
  including the emerging quasinormal modes in such space-times, possible gravitational wave
  echoes and gravitational lensing phenomena \cite{churilova19, yang21, guerrero21, tsukamoto21,
  islam21, cheng21, kb20, tsukamoto20, lima21, nascimento20, franzin22}.
  
  This paper continues the studies of various \ssph\ space-times, including regular ones, starting 
  from the general form of the metric,
\beq             \label{ds}
		ds^2 = A(x) dt^2 - \frac {dx^2}{A(x)} - r^2(x) d\Omega^2,
\eeq  
  written in terms of the so-called quasiglobal radial coordinate $x$ \cite{BR-book}, the choice 
  particularly convenient for the description of \bh\ and \wh\ space-times. Recall that \bh\ 
  horizons correspond to regular zeros of $A(x)$ provided that $r(x)$ is finite, while \wh\ throats 
  are described as regular minima of $r(x)$ provided that $A(x) > 0$. If a minimum of $r(x)$ 
  takes place in a region where $A(x) < 0$, it is called a black bounce \cite{simp18}, while a 
  minimum of $r(x)$ coinciding with a horizon ($A=0$) has been named a black throat \cite{rahul22}. 
		
  The metric \rf{ds}, being substituted to the Einstein equations, leads to the stress-energy
  tensor (SET) of an anisotropic fluid, that is, $T\mN = \diag(\rho, -p_r, - p_\bot, -p_\bot)$, 
  with the density $\rho$, radial pressure $p_r$ and tangential pressure $p_\bot$. 
  It is, however, more desirable to have a field representation 
  of SV-like regular space-times, in which case their Lagrangian formulation makes it possible a
  further study of their dynamics, including their stability properties. In this respect, we can recall 
  that, according to \cite{rahul22, canate22, kb22}, an arbitrary \ssph\ metric can be obtained as 
  an exact solution to Einstein's equations with a SET containing two contributions: one from a 
  self-interacting scalar field minimally coupled to gravity, and another one from a nonlinear \emag\ 
  field with the Lagrangian of the form $L_e = L_e(\cF)$, where $\cF = F\mn F\MN$, $F\mn$ 
  being the \emag\ field tensor. It turns out that a scalar field alone or an \emag\ field alone
  cannot account for an arbitrary metric, including its SV-like examples,  because of the symmetry 
  properties of their SETs: $T^t_t = T^x_x$ for \emag\ fields and $T^t_t = T^\theta_\theta$ 
  for scalar fields. Meanwhile, their combination allows for obtaining a source of any \ssph\ metric
  \cite{kb22}.
      
  In the sources of the metrics containing either \wh\ throats or black bounces (or both), the presence 
  of a phantom field is necessary, since the corresponding SET should violate the Null Energy Condition 
  (NEC). This circumstance is confirmed by the explicit forms of the field 
  sources obtained for SV regularizations of the \Scw\ and \RN\ metrics \cite{rahul22}. However, 
  in the general case it turns out that the scalar field cannot be only canonical (that is, with positive 
  kinetic energy) or only phantom (with negative kinetic energy), but must change its nature when 
  passing from one region to another \cite{kb22}. Such a scalar, which is phantom in a
  strong-field region and becomes canonical elsewhere, was called a ``trapped ghost'' and was 
  used to obtain a number of globally regular space-times including \whs\ and regular \bhs\
  \cite{kroger04, trap10, don11}, in particular, those generalizing regular \wh\ and \bh\
  solutions (``black universes'') obtained with a purely phantom scalar  
  \cite{bu1, bu2, bu3}. Moreover, it was shown that transition surfaces on which 
  a scalar field changes its nature can play a stabilizing role in \bh\ and \wh\ models
  \cite{trap17, trap18}. Examples of SV-like regularizations of Fisher's solution for a 
  canonical massless scalar field in GR \cite{fisher48} and a subset of dilatonic \bh\ solutions
  with interacting massless \emag\ and scalar and fields \cite{dil1, dil2, dil3} have been obtained 
  in \cite{kb20}. In both cases, the corresponding sources contained a nonlinear \emag\ field
  and a trapped-ghost scalar field.
  
  In this paper we try to obtain a field description of an arbitrary \ssph\ space-time with a single 
  field rather than a combination of two fields. It is suggested to use, as a necessary new degree of 
  freedom, a nonminimal coupling between a scalar field and the space-time curvature, in other 
  words, to invoke scalar-tensor theories (STTs) of gravity. We show that such a representation 
  is really possible and describe an algorithm of obtaining the STT characteristics (the nonminimal 
  coupling function $f(\phi)$, the potential $U(\phi)$ and the kinetic coupling function $h(\phi)$) 
  for arbitrary $A(x)$ and $r(x)$ in the metric \rf{ds}. Then we apply this algorithm to two 
  particular examples, including \rf{ds-SV}. An important (though undesired) feature of this 
  algorithm is that it works piecewise, such that different regions of its validity are separated 
  by surfaces where $f =0$ or $f = \infty$.
  
  We adopt the metric signature $(+\,-\,-\,-)$ and the geometrized units such that $8\pi G = c = 1$,
  so that the Einstein equations are used in the form 
  $G\mN \equiv R\mN - \half \delta\mN R = - T\mN$. 
  
\section{STT representation of \sph\ space-times}  

  The general Bergmann-Wagoner-Nordtvedt STT of gravity is described by the action
  \cite{STT1, STT2, STT3}
\bearr   \label{S}
             S_{\rm STT} = \frac 1{16\pi} \int \sqrt{-g} d^4 x
		             \Big[f(\phi) R + 2 h(\phi)\phi^{,\alpha} \phi_{,\alpha} 
\nnn \inch             
  			           - 2 U(\phi) + L_m\Big],
\ear    
  where $R$ is the scalar curvature of space-time, $f, h$, and $U$ are arbitrary functions of
  the scalar field $\phi$ (where $f(\phi)> 0$ should be finite since the effective gravitational 
  constant is proportional to $1/f)$), and $L_m$ is the nongravitational matter Lagrangian. 
  Assuming $L_m=0$, the field equations that follow from \rf{S} have the form
\bearr       \label{EE}
		(G\mN + \nabla_\mu \nabla^\nu - \delta\mN \Box) f
\nnn		\cm
		+ h (2\phi_{\mu}\phi^{,\nu}- \delta\mN \phi^{,\alpha} \phi_{,\alpha}) + \delta\mN U =0,
\yyy             \label{eq-phi}
		4h \Box\phi +2 h_\phi  \phi^{,\alpha} \phi_{,\alpha} - f_\phi R + 2 U_\phi =0,  
\ear      
  where $G\mN$ is the Einstein tensor, $\Box = \nabla^\mu\nabla_\mu$
  is the d'Alembert operator, and the subscript $\phi$ denotes $d/d\phi$. As follows from the 
  contracted Bianchi identity $\nabla_\nu G\mN=0$, \eqn {eq-phi} is a consequence of \rf{EE}, 
  therefore, in what follows we only consider \rf{EE}.  
  
  For the metric \rf{ds}, the Einstein tensor has the following nonzero components:
\bearr           \label{G00}
		G^t_t = \frac 1{r^2} \big[-1 +A(2r r'' + r'{}^2) + A'rr'\big],
\\ \lal           \label{G11}
		G^x_x = \frac 1{r^2} \big[-1 + A' r r' + A r'{}^2\big],
\\ \lal           \label{G11}
		G^\theta_\theta = G^\varphi_\varphi =  \frac 1 r\big[Ar''+ \half A''r + A'r'\big].
\ear  
  Three noncoinciding equations \rf{EE} for the same metric have the form
\bearr          \label{E00}
  		f G^t_t +\! A f'' + f'\Big(\!\frac{2Ar'}{r}\!+ \! \frac{A'}{2}\!\Big) +\! Ah\phi'{}^2\! +U =0, 
\yyy           \label{E11}
  		f G^x_x + f'\Big( \frac{2Ar'}{r} + \frac{A'}{2}\Big) -\! Ah\phi'{}^2\! +U =0,  
\yyy           \label{E22}
  		f G^\theta_\theta +\! A f'' + f'\Big(\!\frac{Ar'}{r}\! +\! A'\Big) + \! Ah\phi'{}^2\! + U =0,
\ear
  where we have assumed $\phi = \phi(x)$. Let us form three combinations of these equations: 
\bearr          \label{E02}
  		f (G^t_t - G^\theta_\theta) = f' \Big(\frac{A'}{2} - \frac{Ar'}{r} \Big), 
\yyy           \label{E12}
  		f (G^t_t + G^x_x) +\!Af''\!+ f'\Big(\frac{4Ar'}{r}\!+\! A'\Big)\! =\! - 2U,  
\yyy           \label{E01}
  		f (G^t_t - G^x_x) + A f'' =-  2 Ah \phi'{}^2.  
\ear

  Now, if the functions $A(x)$ and $r(x)$ are known smooth functions and $r\ne 0$, all components 
  of $G\mN$ are known and smooth as well, and integration of \eqn{E02} gives $f$ as a function 
  of the coordinate $x$. Next, as soon as we know $f(x)$, we can calculate the quantities $U$ from 
  \rf{E12} and $h \phi'{}^2$ as functions of $x$. 
  Since the latter quantity can have a variable sign, it makes sense to choose 
  ``by hand'' the function $\phi(x)$ as any monotonic function, making it easy to convert $f(x)$ 
  and $U(x)$ into funtions of $\phi$. This step does not break the generality but only fixes the scalar 
  field parametrization. Then $h(x)$ and hence $h(\phi)$ are easily determined, and our metric has thus 
  been presented as a vacuum solution of a certain STT.  
  
  A difficulty emerges if $F(x) = A'/2 - Ar'/r = 0$ at some $x=x_0$. Generically, at such points 
  $(G^t_t - G^\theta_\theta)\big|_{x_0} = G_0 \ne 0$.  At its regular zeros, the function $F(x)$ 
  behaves as $k(x-x_0)^n$, where $n \in \N$, $k \in \R$, and $k \ne 0$. Therefore, near $x_0$, 
  we have $\ln f \sim (G_0/k) \int x^{-n} dx$, a divergent integral (logarithmically in the generic 
  case of a simple zero, $n =1$, and in a power manner otherwise). Therefore,  as $x\to x_0$, 
  $f(\phi)$ tends either to zero or to infinity, depending on the sign of the ratio $G_0/k$. 
  We have to conclude that our STT representation fails at such points  $x=x_0$ and is valid only 
  in ranges of $x$ where $F(x) \ne 0$. As a result, the following theorem can be formulated:
  
\medskip\noi
  {\bf Theorem:} {\sl An arbitrary \ssph\ metric in the form \rf{ds}, specified in a range of $x$ such 
  that $F(x) = A'/2 - Ar'/r \ne 0$, is a solution to the vacuum field equations of a certain scalar-tensor
  theory of the form \rf{S} with functions $f(\phi) >0$, $h(\phi)$ and $U(\phi)$ that are regular 
  in this range of $x$.}  
  
\medskip      
  We can notice that possible zeros of $G^t_t - G^\theta_\theta$ make no problem. Even more
  than that: the trivial case $G^t_t - G^\theta_\theta =0$ corresponds to $f = \const$, hence 
  to any solutions of GR sourced by a minimally coupled scalar field $\phi$ with an arbitrary 
  self-interaction potential $U(\phi)$. We also notice that the sign of the function $h(\phi)$
  which, if $f=\const$,  determines the canonical ($h > 0$) or phantom ($h < 0$) nature of 
  the $\phi$ field, is not fixed, and we can in general obtain the situation of the so-called
  ``trapped ghost''  \cite{trap10, don11, kb22}, where $h(\phi)$ changes its sign on some
  transition surface, without leading to a space-time singularity. The sign of $h(\phi)$ in STT 
  is also not fixed, but its relationship with a possible phantom nature of $\phi$ is not so direct 
  and is connected with a transition to the Einstein conformal frame \cite{STT2}. Specifically,
  the scalar field in STT is phantom if 
\beq                 \label{D}
				D(\phi) =2 f(\phi) h(\phi) + \frac 32 \Big(\frac{df}{d\phi}\Big)^2 < 0. 
\eeq  
  (For example, in the Brans-Dicke STT we have $f = \phi$, $2h = \omega/\phi$, and 
  $D(\phi) = \omega +3/2$, where $\omega$ is the Brans-Dicke coupling constant.) It is also
  important to notice that $\sign D(\phi)$ does not depend on the parametrization of $\phi$:
  indeed, at a transition to another field $\psi$ with a monotonic function $\phi(\psi)$, we obtain 
  $D(\psi) = D(\phi) (d\phi/d\psi)^2$. 
      
\section{Examples}  

  It is clear that considering any vacuum solution of GR or solutions with a minimally coupled 
  scalar field $\phi$ as examples of the procedure under consideration is trivial since such 
  a theory is just the simplest example of STT, and there is nothing to seek. For example, a 
  consideration of the \Scw-(anti-)de Sitter metric will lead to nothing else but $f = \const$, 
  $\phi = \const$ and $U = \Lambda$, the cosmological constant. The same can be said about 
  any solutions of known STT, such as the Brans-Dicke theory or GR with a conformally coupled 
  scalar field (see, e.g., \cite{kb73} for general solutions with $U=0$). However, all other 
  assumptions of the metric require a full consideration as described in the previous section. 
  We will consider here two such examples: the \RN\ metric and the SV regularization of the 
  \Scw\ metric. 
  
\subsection{The \RN\ metric}  

  The \RN\ metric has the form \rf{ds} with
\beq                \label{ds-RN}
 		r(x) \equiv  x, \quad\  A(x) = A(r) = 1 - \frac{2m}{r} + \frac{q^2}{r^2}, 
\eeq  
  where $m$ is the \Scw\ mass, and $q$ is the charge which can be electric or magnetic
  or mixed. It is a solution to the Einstein equations whose source is a \sph\ Maxwell 
  electromagnetic field, with the stress-energy tensor of the form 
\beq
		T\mN = \frac{q^2}{r^4} \diag (+1, +1, -1, -1).
\eeq    
  This expression can be directly used in \eqs \rf {E02}--\rf{E01} after the substitution 
  of the Einstein equations $G\mN = - T\mN$. Consequently, \eqs \rf {E02}--\rf{E01} with
  the functions \rf{ds-RN} take the form (the prime denotes $d/dr$)
\bearr                  \label{02-RN}
		\frac{2q^2}{r^4}f = \frac{r^2 - 3mr + 2r^2}{r^3} f',
\\ \lal                \label{01-RN}
		\frac{2q^2}{r^4}f - \frac{r^2 - 2mr+q^2}{r^2} f'' - \frac{4r^2-6mr + 2q^2}{r^3}f'
\nnn  \cm\cm\inch 		
				  = 2U,
\\ \lal		                \label{0-1-RN}
           f'' = -2 h \phi'{}^2.
\ear

  Let us restrict ourselves to the case $9m^2 > 8 q^2$, such that the trinomial $r^2 - 3mr + 2q^2$
  has different real roots $(a, b) = \half(3m \pm \sqrt{9m^2 - 8q^2})$ (otherwise the results for 
  $f(\phi)$ and other quantities will look more complicated).\footnote
  		{Note that $a$ and $b$ do not coincide with the horizon radii $r_\pm$ given by
  		$r_\pm = m \pm \sqrt{m^2 - q^2}$.}  
  We thus have $a+b = 3m$ and $ab = 2q^2$  	
  	
  Integrating \eqn{02-RN}, we obtain
\beq              \label{f-RN}
			f(r) = f_0 r |r-a|^{b/(a-b)} |r-b|^{-a(a-b)}, 
\eeq   
  with an integration constant $f_0 >0$. Next, \eqn{01-RN} yields an expression for the potential
  $U(r)$:
\beq           \label{U-RN}
		U(r) = \frac{f(r)}{2}\, \frac{ab (2a^2 - 5ab + 2b^2)}{6r^2 (r-a)^2 (r-b)^2} ,
\eeq  
  while \eqn{0-1-RN} gives us the kinetic term as
\beq               \label{h-RN}
 		h \phi'{}^2 = f(r) \, \frac{ab(3r - 2a - 2b)}{r(r-a)^2(r-b)^2}
\eeq
    
   Without loss of generality, we can choose the parametrization of the $\phi$ field as 
\beq                     \label{phi-S}
		\phi = \frac 1m \arctan \frac rm \ \ \then \ \ \phi' = \frac 1{r^2+ m^2}. 
\eeq
  Then the quantities $f(\phi)$, $h(\phi)$, $U(\phi)$, being calculated as functions of $r$, are easily 
  transformed to the variable $\phi$ since now $r = m \tan(m\phi)$. Since $ \phi'{}^2 >0$, we can 
  judge on the sign of $h(\phi)$ by the r.h.s. of \eqn{h-RN}. Evidently, it coincides with the sign of 
  $3 r - 2a -2b = 3 (r-2m)$,  so that $h > 0$ at $r > 2m$ and $h <0 $ at $r < 2m$.   
  A calculation of $D$ according to \rf{D} shows that 
  $\sign D = \sign(2r^2 -4mr +q^2)$, hence a ``phantom domain'' occurs 
  at $ m - \sqrt{m^2 - q^2/2} < r < m + \sqrt{m^2 - q^2/2}$.
  
  From \eqn{f-RN} it follows that the STT representations of the \RN\ space-time are different in 
  the ranges $r \in (0, b)$, $r \in (b,a)$, and $r > a$. The boundaries of these ranges are in no way 
  connected with physically distinguished spheres like horizons. 
  
  In the special case $q =m$ (an extremal \RN\ \bh) we have $a=2m$, $b =m$ and $U=0$.
  This metric is well known to be simultaneously a solution to the special case of STT representing 
  \GR\ with a massless conformally coupled scalar field, corresponding to $f(\phi) = 1 - \phi^2/6$
  and $h(\phi) = 1/2$, $U=0$ \cite{kb70, kb73, bek74}. Such results are really obtained in our 
  consideration, with 
\beq                      \label{f-RN0}
		\phi = \frac{\sqrt{6} m}{r -m},\qq       f = \frac {r(r-2m)}{(r-m)^2},
\eeq		   
   thus confirming the correctness of our calculations.
  
   One can also notice that, by \rf{f-RN0}, STT representations of the extreme \RN\ metric take place
   separately in the regions $r > 2m$, $r \in (m, 2m)$, ans $r < m$. For example, a transition to the 
   Einstein frame (or vice versa, since the STT solution is easier obtained in the Einstein frame) 
   can be implemented by a conformal mapping with the factor $f(\phi)$ in the range $r > 2m$, 
   and after that the full STT solution is obtained by a further extension to $r \leq 2m$,
   being an example of conformal continuation \cite{kb-conf}, an operation that is necessary when 
   the Einstein-frame space-time maps to only a region of the Jordan-frame space-time.

\subsection{SV regularization of the \Scw\ metric}

   Now let us consider the metric \rf{ds-SV}, so that in \rf{ds} we have
\beq                \label{ds-SV+}
 		r(x) =\sqrt{x^2 + b^2}, \quad\  A(x) = 1 - \frac {2m}{\sqrt{x^2 + b^2}}
\eeq  
  where $m$ has the meaning of the \Scw\ mass, and $b>0$ is the regularization parameter
  With the functions \rf{ds-SV+}, \eqs \rf {E02}--\rf{E01} take the form
\bearr             \label{02S}
			x (r - 3m) f' = \frac{3mb^2}{r^2} f,
\yyy	             \label{01S}		
			\frac{r\! - \! 2m}{r} f'' + \frac {2x}{r^3}(2r\! - \! 3m)f' - \frac{4mb^2}{r^5} f =\! -2U, 
\yyy             \label{0-1S}
			f'' + \frac{2b^4}{r^4} f = - 2 h \phi'{}^2,
\ear
  where $r = r(x)$ as given in \rf{ds-SV+}, and the prime denotes $d/dx$.
  
  Integration of \rf{02S} gives, for $b \ne 3m$, 
\bearr                     \label{f-S}
 			f(x) = f_0 r (r+b)^{-\tfrac{3m}{2(3m +b)}} 
 						  (r- b)^{-\tfrac{3m}{2(3m -b)}} 
\nnn \cm\times 						  
 						  |r-3m|^{\tfrac{b^2}{9m^2 -b^2}},  \qq  f_0 = \const,
\ear
  and for $b = 3m$,
\beq                     \label{f-S3}
                f(x) = f_0 r (r+b)^{-1/4} (r-b)^{-3/4} \e^{-b/(2r-2b)}.   
\eeq   

  As before, without loss of generality, we can choose the parametrization of the $\phi$ field as 
\beq                     \label{phi-S}
		\phi = \frac 1b \arctan \frac xb \ \ \then \ \ \phi' = \frac 1{x^2+b^2} = \frac 1{r^2}.
\eeq
  Then the quantities $f(\phi)$, $h(\phi)$, $U(\phi)$, being calculated as functions of $x$ or
  $r$, are easily transformed to the variable $\phi$ since now $x = b \tan(b\phi)$ and 
  $r = b/\cos (b\phi)$.
  		
  Consider the case of more interest from the viewpoint of regularization< when the parameter 
  $b$ is small, $b < 3m$. From \rf{0-1S} we obtain an expression for $ h \phi'{}^2$, and, with 
  the ansatz \rf{phi-S},
\bearr  
  			h(\phi) = \frac{b^2 f(\phi)}{2}
\nnn \ \ \times  			
  			\bigg[\frac {3m[4r^3\! -\!9mr^2+ 3b^2(m\!-\! r)]}
				{(r^2 -b^2)(r-3m)^2}- 2\bigg],
\ear  
  with $f(\phi)$ given by \rf{f-S}. The potential $U(\phi)$ is found in a similar way from \rf{01S},
  with the result 
\bearr  
			U(\phi) = - \frac{m b^2 f(\phi)}{2}
\nnn \ \ \times			
			\frac{b^2 (3 m + r) + r (-36 m^2 + 21 m r - 4 r^2)}	{r^4 (r^2-b^2) (r-3 m)^2}.
\ear

  The function \rf{f-S} tends to infinity as $r\to b$ (that is, at the black bounce $x=0$ and to
  zero at $r = 3m$, so that we have as many as four separate regions of STT representation:
\[  \nq\,
         x < - x_0, \quad  -x_0 < x < 0, \quad  0 < x < x_0, \quad   x > x_0, 
\]   
  where $x_0 = \sqrt{9m^2 - b^2}$. The behavior of the functions $h(\phi)$ and $U(\phi)$ 
  is the same at both positive and negative $x$ and is illustrated in Fig.\,1 in terms of the radius $r$.

  It is clear that the metric \rf{ds-SV+} containing a black bounce requires a phantom behavior
  of its source. However, curiously, in our STT representation $h(\phi)$ still becomes positive at 
  sufficiently large $|x|$ or $r$, thus demonstrating ``trapped-ghost'' properties.
  
\begin{figure*}
\centering
\includegraphics[width=7.5cm]{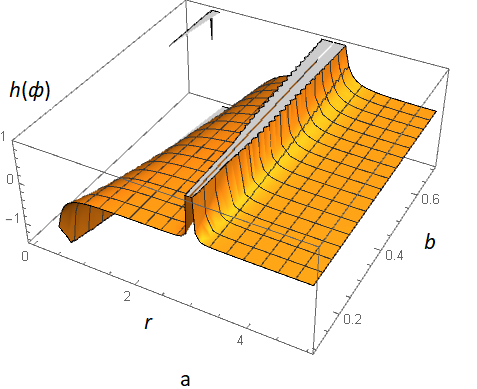} \ \ \
\includegraphics[width=7.5cm]{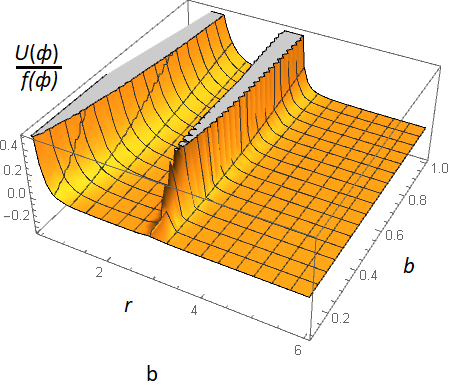}
\caption{\small
      The quantities $h(\phi)$ (panel a) and $U(\phi)/f(\phi)$ (panel b) as functions of the radius 
      $r$ and the regulrization parameter $b$, assuming $m=1$. The peak at $r = 3m =3$
      separates different regions of the STT representation.}
\bigskip      
\end{figure*} 		
  
\section{Concluding remarks}

  We have shown that any \ssph\ metric can be presented as a vacuum solution to the equations of 
  some STT \rf{S} in its Jordan frame. An evident exception is the case of $A = \const\cdot r^2$
  that implies $F(x) = A'/2 - Ar'/r = 0$ making it impossible to determine $f(\phi)$. As we saw,
  there are in general some values of the radial coordinate $x$ where $F=0$, and such values 
  separate regions of different STT representations. It also follows that the corresponding 
  Einstein-frame metrics, obtained from an original metric under consideration with the conformal
  factor $f(\phi)$, will be in general singular where $f$ is zero or infinite, hence the whole 
  Einstein-frame space-time maps to only a region of the Jordan-frame space-time, making 
  conformal continuations \cite{kb-conf} quite a generic phenomenon in STT representations of
  different metrics.    
  
  Another generic phenomenon is the emergence of separate regions where the kinetic coefficient 
  $h(\phi)$ is negative, leading to the so-called trapped ghosts \cite{trap10, don11, trap17, trap18}. 
  It is of interest that a ``ghost indicator'' $D <0$ is observed even in some range of radii in \RN\ 
  space-time despite the absence of any exotic source of this metric in GR. On the other hand, in the
  case of the metric \rf{ds-SV}, where a phantom source is evidently necessary, we obtain a canonical
  behavior of the scalar field outside a small strong-field region.
  
  A shortcoming of our STT representation is that the separating values of $x$ where $F=0$ are
  in general not clearly related to physically distinguished surfaces of the space-time under study.
  Nevertheless, the very existence of this representation as well as its properties look instructive and 
  potentially useful.
  
\bigskip  
  
\Acknow{K.B. gratefully acknowledges the outstanding hospitality of the colleagues from Samarkand State 
	University, where  this study was initiated.} 
   
\subsection*{Funding}

   K.B. acknowledges partial support from the RUDN University Strategic Academic Leadership 
   Program,  and from the Ministry of Science and Higher Education of the Russian Federation, Project
   ``Fundamental properties of elementary particles and cosmology" N 0723-2020-0041. 
   K. Badalov and  R. Ibadov  gratefully acknowledge the support from Ministry of Innovative 
   Development of the Republic of Uzbekistan, Project  No. FZ-20200929385.     

\ConflictThey    
  
\newpage

\end{document}

%% file: preamble.tex
\usepackage{amsfonts,amssymb,cite}
\usepackage{graphicx}

\def\noi{\noindent}

\newcommand{\Title}[1]{\noi {{\Large\bf #1}}\\[1ex]}

\def\Aunames#1{\noi{\bf #1}}
\def\au#1{${}^{#1}$}
\def\Addresses#1{\medskip\noi \protect
	\begin{description}\itemsep -3pt {\it #1} \end{description}}
\def\adr#1#2{\item[${}^{#1}$]{\it #2}}

\newcommand{\Abstract}[1]{\vskip 2mm \begin{center}
        \parbox{16.4cm}{\small\noi #1} \end{center}\medskip}

\def\email#1#2{\footnotetext[#1]{e-mail: #2}\addtocounter{footnote}{1}}

\def\nq{\hspace*{-1em}}
\def\nqq{\hspace*{-2em}}

\def\qq{\qquad}
\def\cm{\hspace*{1cm}}
\def\inch{\hspace*{1in}}

\usepackage{color}

\def\Acknow#1{\subsection*{Acknowledgments} #1}

\def\ConflictThey{\subsection*{Conflict of interest} 
	The authors declare that they have no conflicts of interest.}

\def\Jl#1#2{#1 {\bf #2},\ }

\def\ApJ#1 {\Jl{Astroph. J.}{#1}}
\def\CQG#1 {\Jl{Class. Quantum Grav.}{#1}}
\def\DAN#1 {\Jl{Dokl. AN SSSR}{#1}}
\def\GC#1 {\Jl{Grav. Cosmol.}{#1}}
\def\GRG#1 {\Jl{Gen. Rel. Grav.}{#1}}
\def\IJMPD#1 {\Jl{Int. J. Mod. Phys. D}{#1}}
\def\JETF#1 {\Jl{Zh. Eksp. Teor. Fiz.}{#1}}
\def\JETP#1 {\Jl{Sov. Phys. JETP}{#1}}
\def\JHEP#1 {\Jl{JHEP}{#1}}
\def\JMP#1 {\Jl{J. Math. Phys.}{#1}}
\def\NPB#1 {\Jl{Nucl. Phys. B}{#1}}
\def\NP#1 {\Jl{Nucl. Phys.}{#1}}
\def\PLA#1 {\Jl{Phys. Lett. A}{#1}}
\def\PLB#1 {\Jl{Phys. Lett. B}{#1}}
\def\PRD#1 {\Jl{Phys. Rev. D}{#1}}
\def\PRL#1 {\Jl{Phys. Rev. Lett.}{#1}}

\def\lal{&&\nqq {}}
\def\eq{Eq.\,}
\def\eqs{Eqs.\,}
\def\beq{\begin{equation}}
\def\eeq{\end{equation}}
\def\bear{\begin{eqnarray}}
\def\bearr{\begin{eqnarray} \lal}
\def\ear{\end{eqnarray}}
\def\earn{\nonumber \end{eqnarray}}

\def\nnn{\nonumber\\ \lal }
\def\nnnv{\nonumber\\[5pt] \lal }
\def\yy{\\[5pt] {}}
\def\yyy{\\[5pt] \lal }

\def\tst{\textstyle}

\def\fract#1#2{{\tst\frac{#1}{#2}}}

\def\half{{\fract{1}{2}}}

\def\e{{\,\rm e}}

\def\sign{\mathop{\rm sign}\nolimits}
\def\diag{\mathop{\rm diag}\nolimits}

\def\const{{\rm const}}

\def\then{\ \Rightarrow\ }

